\documentstyle[psfig,12pt,aaspp4]{article} % my preference
\def\kms{~km~s$^{-1}$}

\def\be{\begin{equation}}
\def\ee{\end{equation}}

\def\m{~$\mu$m}
\def\HI{\ion{H}{1}}
\def\HII{\ion{H}{2}}

\def\ISO{{\it ISO}}
\def\IRAS{{\it IRAS}}

\def\msun{~$M_\odot$}
\def\sbs{SBS~0335-052}

\begin {document}
%\hskip 3.5in{\scriptsize \today \hskip 0.2in Version 4.1}
\title {Multiwavelength Observations of the Low Metallicity Blue Compact Dwarf Galaxy SBS 0335-052\altaffilmark{1}}
 
\author {Daniel A. Dale,\altaffilmark{2} George Helou,\altaffilmark{3,4} Gerry Neugebauer,\altaffilmark{3} B.T. Soifer,\altaffilmark{3,4} David T. Frayer,\altaffilmark{4} James J. Condon\altaffilmark{5}}
\altaffiltext{1}{\scriptsize Based in part on observations obtained at the W.M. Keck Observatory  which is operated as a scientific partnership among the California Institute of Technology, the University of California and the National Aeronautics and Space Administration.}
\altaffiltext{2}{\scriptsize Division of Physics, Mathematics and Astronomy, California Institute of Technology 320-47, Pasadena, CA 91125}
\altaffiltext{3}{\scriptsize Palomar Observatory, California Institute of Technology 320-47, Pasadena, CA 91125}
\altaffiltext{4}{\scriptsize SIRTF Science Center, California Institute of Technology 314-6, Pasadena, CA 91125}
\altaffiltext{5}{\scriptsize National Radio Astronomy Observatory, 520 Edgemont Road, Charlottesville, VA 22903}
\altaffiltext{6}{\scriptsize The National Radio Astronomy Observatory is a facility of the National Science Foundation operated under cooperative agreement by Associated Universities, Inc.}

\begin{abstract}
New infrared and millimeter observations from Keck, Palomar, \ISO, and OVRO and archival data from the NRAO\altaffilmark{6} VLA and \IRAS\ are presented for the low metallicity blue compact dwarf galaxy \sbs.  Mid-infrared imaging shows this young star-forming system is compact (0\farcs31; 80~pc) at 12.5\m.  The large Br$\gamma$ equivalent width (235~\AA) measured from integral field spectroscopy is indicative of a $\sim$5~Myr starburst.  The central source appears to be optically thin in emission, containing both a warm ($\sim80$~K) and a hot ($\sim210$~K) dust component, and the overall interstellar radiation field is quite intense, about $10,000$ times the intensity in the solar neighborhood.  CO emission is not detected, though the galaxy shows an extremely high global \HI\ gas-to-dust mass ratio, high even for blue compact dwarfs.  Finally, the galaxy's mid-infrared-to-optical and mid-to-near-infrared luminosity ratios are quite high, whereas its far-infrared-to-radio and far-infrared-to-optical flux ratios are surprisingly similar to what is seen in normal star-forming galaxies.  The relatively high {\it bolometric} infrared-to-radio ratio is more easily understood in the context of such a young system with negligible nonthermal radio continuum emission.  These new lines of evidence may outline features common to primordial galaxies found at high redshift.
\end{abstract}
 

\section {Introduction}

The blue compact dwarf galaxy \sbs\ is an extremely metal-poor galaxy ($Z_\odot/41$; Izotov et al. 1997), with only I~Zw~18 known to have a lower metal abundance.  At a radial velocity c$z=4043$\kms, \sbs\ is at a distance of 53~Mpc for $H_{\rm o}=75~{\rm km~s}^{-1}~{\rm Mpc}^{-1}$ (1\arcsec\ corresponds to 258~pc).  Recent observations have highlighted several interesting characteristics of \sbs\ and its fainter companion 80\arcsec\ (22~kpc) to the west.  In addition to showing very low metallicity, the bright starburst region of \sbs\ is situated in an extensive \HI\ envelope, has an extremely large Br$\gamma$ equivalent width, and appears unexpectedly bright in the mid-infrared, with a 12\m-to-$B$-band luminosity ratio greater than two (Pustilnik et al. 2001; Vanzi et al. 2000; Thuan, Sauvage \& Madden 1999).  The last property is particularly intriguing since the galaxy exhibits anemic levels of metals.  Moreover, Thuan, Sauvage \& Madden found that the mid-infrared spectrum lacks the standard mid-infrared dust emission features, is dominated by a strong continuum, and peaks at a surprisingly short wavelength, suggestive that the radiating dust is embedded in an extremely intense interstellar radiation field.  $HST$ optical observations by Papaderos et al. (1998) indicate an underlying stellar population of mass $\sim 3 \times 10^6~M_\odot$ and an age younger than 100~Myr.  Near-infrared data on the handful of bright super star clusters that dominate the galaxy's optical luminosity suggest that the central star-forming region is even younger, of order 5~Myr (Vanzi et al. 2000).

The unusual mid-infrared properties of \sbs\ merit high spatial resolution studies: is the opacity at mid-infrared wavelengths consistent with $A_V \sim 20$~mag (Thuan, Sauvage \& Madden 1999) or $A_V \sim 0.6$~mag (Izotov et al. 1997)?  Can the emission source be resolved?  What fraction of the mid-infrared light is associated with the brightest star clusters?  For example, Soifer et al. (2000) and Soifer et al. (2001) find that nearly all of the luminous and ultraluminous infrared galaxies in their samples emit a substantial fraction ($> 50\%$) of their mid-infrared luminosity in compact $<100-300$~pc-sized cores.  Similar information on dwarf galaxies in general, and \sbs\ in particular, does not exist; mid-infrared data from \IRAS\ and \ISO\ have an angular resolution of several to tens of arcseconds, precluding detailed studies of compact mid-infrared emission.  In this paper we present diffraction limited mid-infrared imaging using the Keck telescope to investigate the innermost regions of \sbs.  Additional near-infrared integral field spectroscopy, and far-infrared, radio and millimeter data are presented to further constrain the energetics of the interstellar medium.  

While the physics of the interstellar medium in low-metallicity dwarf galaxies is interesting in its own right, this relatively nearby system allows us to study in detail a truly young galaxy and thus could be a useful template for understanding galaxy formation at high redshifts.  This comparison would be especially relevant if normal galaxies are assembled from a collection of small systems (e.g. van Dokkum et al. 1999; Cole et al. 2000).  For example, 25\m\ observations with $NGST$ of $z\approx2$ galaxies, some of which may be in their earliest stages of development, may be usefully compared to (similar rest wavelength) mid-infrared observations of \sbs. 
%cz=4043 km/s 03h37m44.0s-05d02m38s (J2000).  In local group rest frame get D=53.3 Mpc assuming Ho=75.  implies 1"->258 pc
%Is SBS @ z=4 observable with MIPS @ 70 micron?  Well, D_L=36021 Mpc 
%and fnu(12.5)_z=4 is 35mJy*(53.3/36021)^2=7.66e-5 mJy~0.1 muJy which 
%is incredibly smaller than expected 1sigma confusion limit of few 
%hundred muJy
%Is SBS @ z=1 observable with MIPS @ 24 micron?  Well, D_L=5872.2 Mpc
%and fnu(12.5)_z=4 is 35mJy*(53.3/5872.2)^2=2.88e-3 mJy~3 muJy which 
%is smaller than SIRTF expected 1sigma confusion limit of 10-20 muJy
%but observable by NGST, where 3 muJy is S/N=10 (R=3; t=1e5s; 25 mu)
%Is SBS @ z=2 observable with MIPS @ 24 micron?  Well, D_L=14242 Mpc
%and fnu(7.7)_z=4 is 8.9mJy*(53.3/14242)^2=1.25e-4 mJy~0.1 muJy which 
%is a lot smaller than SIRTF expected 1sigma confusion limit of ? muJy
%but observable by NGST, where 0.1 muJy is S/N=1 (R=3; t=1e5s; 23 mu)
%http://www.ngst.stsci.edu/sensitivity/sensitivity/sensitivity.html

\section {Observations and Data Reduction}
\subsection {Keck Near- and Mid-Infrared Imaging}
The mid-infrared observations of \sbs\ were made using the imaging mode of the Long Wavelength Spectrograph (LWS; Jones et al. 1993) at the $f$/25 forward Cassegrain focus of the Keck I Telescope.  The object was centered in the LWS field of view by first imaging in $K$-band with the Near-InfraRed Camera (NIRC; Matthews \& Soifer 1994), and then offsetting the position of the brightest peak at $K$ to the center of the LWS field.

The principal observations were made on the night of 7~December~2000 although observations of variable quality were made on 24~November~1999, 26~January~2000, and 10~September~2000.  The December and September runs were photometric.  A chopper was set to a north-south chopping amplitude of 5\arcsec\ at a frequency $\sim$5~Hz.  The telescope was nodded 5\arcsec\ in the north-south direction.  The data were reduced by differencing the two images obtained within the chop pairs at each nod location, and then coadding the resulting positive images.  Because of the chopper and telescope nod spacings employed for the observations, the flux outside a 5\arcsec\ diameter region is suppressed.  The pixel scale is 0\farcs08~pixel$^{-1}$ for the mid-infrared data and 0\farcs15~pixel$^{-1}$ for the near-infrared data.  No flat-field corrections were performed.

Photometric data were obtained at 12.5\m\ ($\Delta \lambda=1.0$\m).  The photometry was calibrated based on observations of the bright stars HR~1457 ($\alpha$~Tau; [12.5~\micron] $=-3.07$~mag) and HR~8775 ($\beta$ Peg; [12.5~\micron] $=-2.55$~mag) whose magnitudes, in turn, were based on \IRAS\ photometry and previous Keck mid-infrared photometry.  The systematic uncertainties in the photometry, based on the internal consistency of the observations, is estimated to be 5\%.  However, the absolute calibration of standard star magnitudes and the photometric zero points each night limit the overall photometric accuracy to 15-20\%.  The flux density corresponding to 0.0~mag was taken as 26.2~Jy, following the prescription given in the Explanatory Supplement to the \IRAS\ Catalogs and Atlases (Beichman et al. 1985).

On the night of 7~December~2000, the total mid-infrared on-target observation time was 25~minutes.  The observations of \sbs\ were interleaved with six observations of the star IRC-10~046 in order to determine a contemporaneous point spread function (PSF).

\subsection {Palomar Near-Infrared Integral Field Spectroscopy}
Near-infrared observations of the Br$\gamma$ line were carried out on the night of 14~October~2000 using the Palomar Integral Field Spectrograph (PIFS; Murphy, Matthews \& Soifer 1999) on the Palomar 200-inch Telescope.  PIFS provides a 5\farcs4$\times$9\farcs6 field of view using eight separate 0\farcs67$\times$9\farcs6 slits to feed two independent spectrographs within the same liquid N$_2$-cooled dewar.  Each slit is four pixels wide and 58 pixels long (0\farcs167~pixel$^{-1}$).  The high resolution mode was used, providing a spectral resolution of $\lambda/\Delta \lambda \sim 1450$ ($\Delta v \sim 200$\kms) near the observed center wavelength of 2.195\m.

The PIFS observations consisted of a set of eight five-minute on-off integrations.  Thus 40 minutes were spent integrating on source, and an equal amount of time off source (using a 30\arcsec\ nod to the north).  For these observations, the 5\farcs4$\times$9\farcs6 field of view was oriented with the long axis at a position angle of 90\arcdeg\ (the effect of differential atmospheric refraction is minimal for integral field spectroscopy).  A positional dithering pattern was employed for the sequence of observations to recover seeing-limited spatial sampling in the cross-slit direction.  Spectral calibration lamp data were taken immediately afterwards; wavelength calibration is provided through a combination of these data and the available OH airglow lines (Oliva \& Origlia 1992).  The observations were taken under photometric conditions.  Spectrophotometric calibration was performed via spectral imaging of the Persson et al. (1998) standard star 9101, thereby establishing the continuum flux density at 2.195\m; the 2.195\m\ flux density of the standard star was estimated from an extrapolation of its $Ks$-band ($\lambda_{\rm c}=2.157$\m) and $K$-band ($\lambda_{\rm c}=2.179$\m) flux density trend.  Corrections for atmospheric opacity and spectral flat-fielding are derived from observations of HR 1232, a 5.8 magnitude G9V star.  Details of the calibration technique and the PIFS data reduction process are given in Murphy et al. (2000).

An estimate of the near-infrared PSF was obtained after the first 20 minutes of on-source integration and after the second 20 minutes of on-source integration, with each measurement comprised of four dithered ten-second exposures in $K$ band of a nearby field star using the PIFS imaging mode.   An azimuthally averaged radial profile of the PSF target in the first case yields a full width at half maximum (FWHM) of approximately 0\farcs8.  After refocusing, our second measurement was FWHM$\approx$0\farcs5, indicating that the observations were performed in generally good seeing conditions.

\subsection {Owens Valley CO Maps}
\sbs\ was observed using the Owens Valley Millimeter Array between February and May of 1999.  The source was observed in two configurations of the six 10.4~m telescopes with baselines ranging from 15~m to 119~m.  The lines $^{12}$CO($J=1\rightarrow0$) and $^{12}$CO($J=2\rightarrow1$) were simultaneously observed assuming a redshift of 4076\kms\ (Thuan, Izotov \& Lipovetsky 1997).  The CO(1-0) data were recorded using a digital correlator configured with 64$\times$2~MHz channels centered on 113.2750~GHz, while the CO(2-1) line was observed with 59$\times$4~MHz channels centered on 227.4456~GHz.  Both spectrometer setups provided approximately 300\kms\ of bandwidth coverage, which is more than adequate for \sbs.  In addition to the spectral line data, the 3~mm continuum and 1~mm continuum were observed with a 1~GHz bandwidth for both the lower and upper sidebands.  Typical single-sideband system temperatures were approximately 800~K, corrected for telescope losses and the atmosphere.  The nearby quasar PKS~B$0420-014$ was observed every 20 minutes for gain and phase calibration.  Absolute flux calibration was determined from observations of Uranus, Neptune, and 3C\,273.  The absolute calibration uncertainty for the data is approximately 15\% at 3~mm and 25\% at 1~mm.  After data editing, we obtained effectively 20.6 hours of on-source integration time at 3~mm and 6.6 hours of good integration time at 1~mm.  The beam sizes computed using natural weighting to minimize the noise are 5\farcs$4\times$3\farcs8 and 2\farcs9$\times$2\farcs0 for the 113.2750~GHz and the 227.4456~GHz maps, respectively, providing spatial resolution of 1.4~kpc $\times$ 1.0~kpc and 750~pc $\times$ 520~pc.

\subsection {Very Large Array Radio Continuum Maps}
VLA archival data are available at 1.49~GHz from the B configuration.  The data processing includes the primary-beam correction and results in an image with a 6\arcsec\ FWHM Gaussian beam and an rms noise level of 0.13~mJy~beam$^{-1}$.  The point source sensitivity is about a factor of three better than the NRAO VLA Sky Survey (NVSS; Condon et al. 1998), but the surface brightness sensitivity is about 20 times poorer (the NVSS has a 45\arcsec\ FWHM beam).

\subsection {Infrared Space Observatory Far-Infrared Photometry}
\label{sec:ISO}
Results from ISOPHOT observations were kindly provided by M. Sauvage.  \sbs\ was detected with the PHT-C C100 camera system and the 50\m\ filter ($\lambda_{\rm ref}=65$\m, $\lambda_{\rm cen} = 67.3$\m, $\Delta \lambda = 57.8$\m), with a far-infrared flux density of $f_\nu(67$\m)=112~mJy for $S/N \approx 4$.  

\subsection {Data from the Infrared Astronomical Satellite}
\label{sec:IRAS}
\sbs\ was not detected by \IRAS, but robust estimates of the 12, 25, 60 and 100\m\ rms levels can be estimated from the (seven to ten) separate scans made by \IRAS\ at this location.  The SCANPI software has been used to determine the median rms levels.

\section{Results}

\subsection{Mid-Infrared Properties}

\begin{figure}[!ht]
\centerline{\psfig{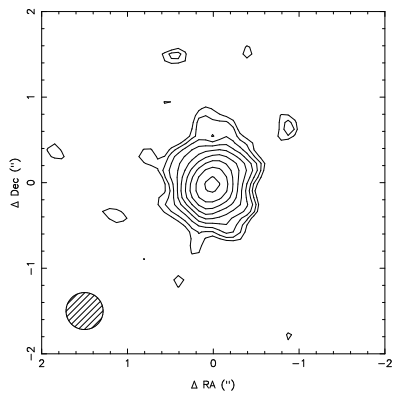}}
\caption[]
{\ Keck I 12.5\m\ image of \sbs, taken on the night of 7 December 2000.  The contours are scaled by factors of 1.342, with the highest and third-highest contours enclosing regions where the emission is greater than 90\% and 50\% of the peak flux, respectively; the lowest contour is at the 3$\sigma$ level.  The FWHM of the PSF is displayed in the lower lefthand corner.  North is up, east is to the left, and the central coordinates are those given in Table~1.}
\label{fig:sbs0335}
\end{figure}

The 12.5\m\ image of \sbs\ is presented in Figure~\ref{fig:sbs0335}.  From the Keck observations the 12.5\m\ magnitude of \sbs\ is 7.2$\pm$0.2~mags which corresponds to 35$\pm$6~mJy (6\% statistical and 15\% systematic uncertainties).  The statistical uncertainty is the standard deviation based on six differences from the average value.  This flux density over a bandpass spanning approximately 12.0 to 13.0\m\ is consistent with the \ISO\ mid-infrared data obtained by Thuan, Sauvage \& Madden (1999), from which they estimate $f_\nu(12$\m)$\simeq21\pm6$~mJy and $f_\nu(14.9$\m)$\simeq35\pm11$~mJy from ISOCAM photometry using the LW10 ($\Delta \lambda = 7.0$\m) and LW9 ($\Delta \lambda = 2.0$\m) filters, respectively (30\% uncertainties have been attached to the \ISO\ fluxes, reflecting the typical calibration uncertainty in such data).  Moreover, the ISOCAM CVF spectrum suggests a 12.5\m\ flux level near 30~mJy.  At a distance of 53~Mpc, the mid-infrared luminosity of \sbs\ is $\nu L_\nu(12.5~\micron) \equiv L(12.5~\micron) = 7.5 \times 10^8~L_\odot$ and the far-infrared luminosity is $L(67~\micron) = 4.4 \times 10^8~L_\odot$.
% where $L \equiv 4 \pi d^2 \nu f_\nu$ and $L_\odot=3.826 \times 10^{26}$~W is taken to be the solar bolometric luminosity.
Table~1 summarizes the main results and general properties of \sbs.

As pointed out by Thuan, Sauvage \& Madden (1999), the \IRAS-equivalent 12\m-to-blue ratio for \sbs\ is several times greater than found for spiral galaxies and more than twice as large as the typical value for blue compact dwarf galaxies.  The mean ratio for the Thuan \& Martin (1981) catalog of 115 blue compact dwarfs is $\left<\log L(12~\micron)/L(0.44~\micron)\right>=-0.12\pm0.27$, and $\left<\log L(12~\micron)/L(0.44~\micron)\right>=-0.55\pm0.06$ for the 69 normal star-forming galaxies described in Helou et al. (1996) and Dale et al. (2000), whereas \sbs\ shows $\log L(12.5~\micron)/L(0.44~\micron) \sim 0.25$.  The mid- to near-infrared and mid-to-far-infrared luminosity ratios are likewise elevated: $\log L(12.5~\micron)/L(2.2~\micron)\simeq1.2$ and $\log L(12.5~\micron)/L(67~\micron)\simeq0.23$, high compared with the average values $\left<\log L(12~\micron)/L(2.2~\micron)\right>=-0.16\pm0.06$ and $\left<\log L(12~\micron)/L(60~\micron)\right>=-0.63\pm0.04$ for normal star-forming galaxies (Dale et al. 2000).  Thus in spite of the galaxy's depressed metallicity, hot dust is one of the energetically dominant features of its interstellar medium.
% assuming L_sun_bolometric=3.826e26 W and d=53.3 Mpc:
%L_2.2 = 3e14/2.2mu*6.32e-24*10**(-k/2.5) =5.099e7 L_sun
%L_12.5 = 35e-29*c/12.5mu*4*pi*d^2 where d=53.3e6*3.09e26 m =7.478e8 L_sun
%L_67.3 = 112e-29*c/67.3mu*4*pi*d^2 where d=53.3e6*3.09e26 m =4.44e8 L_sun
%L_0.44 = 4.26e-23*10^(-0.4*16.96)*c/0.44mu*4*pi*d^2 = 4.252e8 L_sun
%but corrected for galactic extinction, L_0.44 = 5.129e8 L_sun
%see /ull/iso_ground/data/computeB... no internal extinction correction to a
% face-on perspective is applied since assumed it's circular.

The 12.5\m\ source is compact (see Figure~\ref{fig:sbs0335}), and the morphology at this wavelength is consistent with a circularly symmetric distribution of warm dust.  The average observed FWHM of \sbs\ was 0\farcs48$\pm$0\farcs02, whereas the average FWHM of seven separate PSF observations was 0\farcs37$\pm$0\farcs01.  Thus we conclude that \sbs\ is slightly resolved; the intrinsic FWHM after deconvolution of the individual PSFs is 0\farcs31$\pm$0\farcs02 ($80\pm5$~pc).  The uncertainty is statistical and is based on six differences from the FWHM of the PSF.  A Richardson-Lucy deconvolution (Lucy 1974) gives a slightly smaller intrinsic size and likewise shows no significant departures from azimuthal symmetry.  The Keck Telescope and LWS do not provide absolute positions to an accuracy $<$1\arcsec\ and so two independent measurements were made to give the position of the 12.5\m\ image relative to the images at other wavelengths.  NIRC shares a common focal plane with LWS and it is possible to accurately move the telescope from NIRC to LWS.  At 2.2\m\ \sbs\ is dominated by a bright nucleus; measurements of \sbs\ at 12.5\m\ were thus made by offsetting the telescope from this bright nucleus as seen at 2.2\m\ to the LWS array.  To the $\sim\pm$0\farcs5 accuracy of these offsets, the compact source at 12.5\m\ coincides with the bright peak at 2.2\m.  The second relative position measurement is described below.

\subsection{Near-Infrared Integral Field Spectral Data}
\label{sec:NIR_data}
%One of the more intriguing aspects of \sbs\ is that the centroids of the optical continuum and the \HeII\ (4686~\AA) emission appear displaced to the northwest by some 200~pc ($\sim$0\farcs8) with respect to the \hal, H$\beta$, \OIII\ (4363~\AA), \OIII\ (5007~\AA), \HeI\ (5876~\AA), H$_2$ (2.121\m) and Br$\gamma$ (2.166\m) emission lines (Izotov et al. 1997; Vanzi et al. 2000).  The near-infrared continuum in turn is shifted by about 60~pc (0\farcs23) to the northwest of the emission line peaks (Vanzi et al. 2000).  It is unclear whether these discrepancies are due to extinction or real spatial variations for the different emission mechanisms.  Partly because previous spectral information relies only on 1\arcsec-wide long-slit data, integral field spectral data were obtained to better study the properties of the Br$\gamma$ emission and the near-infrared continuum.

\begin{figure}[!ht]
\centerline{\psfig{figure=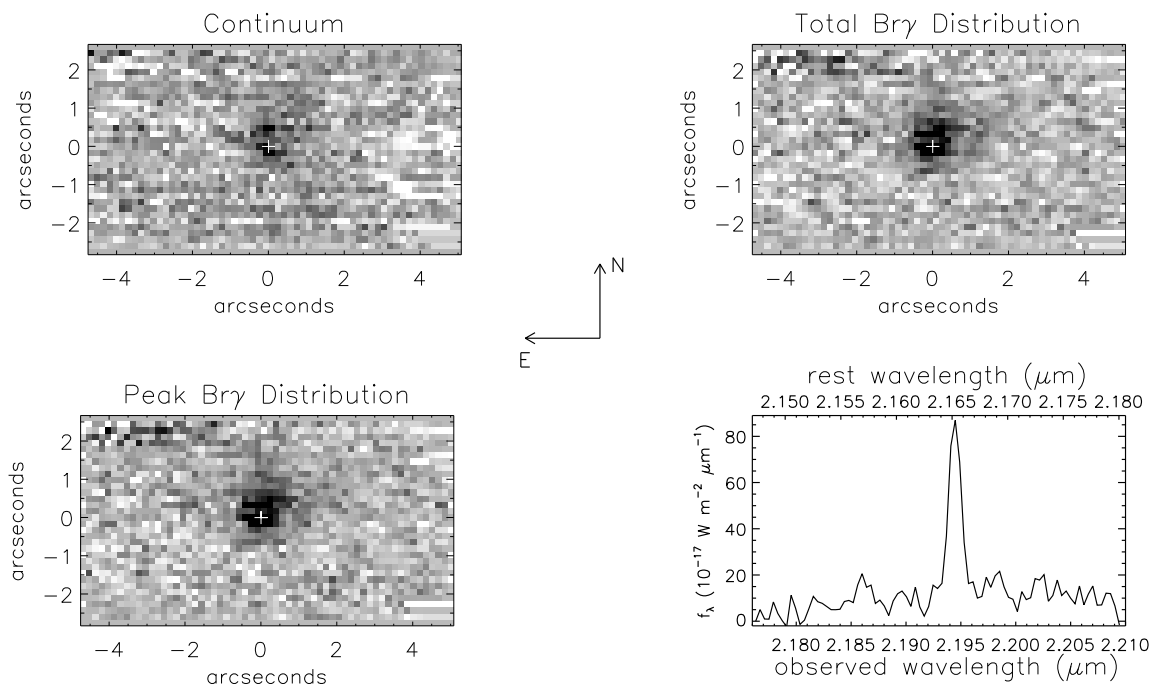,width=6in,bbllx=30pt,bblly=423pt,bburx=585pt,bbury=752pt}}
\caption[]
{\ PIFS Br$\gamma$ data for \sbs.  At top left is the line-free continuum image reconstructed from the datacube.  At top right is the total Br$\gamma$ emission, continuum subtracted, summed from $-170$ to $+170$\kms\ relative to the systemic velocity.  A slightly higher contrast image is shown at lower left, constructed from a single spectral resolution element centered for each spatial pixel on the wavelength of maximum line emission intensity.  In all images, the continuum peak location is indicated by a cross and the central coordinates are those given in Table 1.  Finally, the near-infrared spectrum containing the Br$\gamma$ line is displayed at lower right.}
%lambda_rest=2.1656 lambda_center=2.194805318 
%summed total Brgamma from row 62 to 68 and 0.000404134 microns/pix
%->2.193592916 to 2.19601772 microns, or 3875.2 to 4210.8 km/s
\label{fig:sbs0335_brgamma}
\end{figure}

\begin{figure}[!ht]
\centerline{\psfig{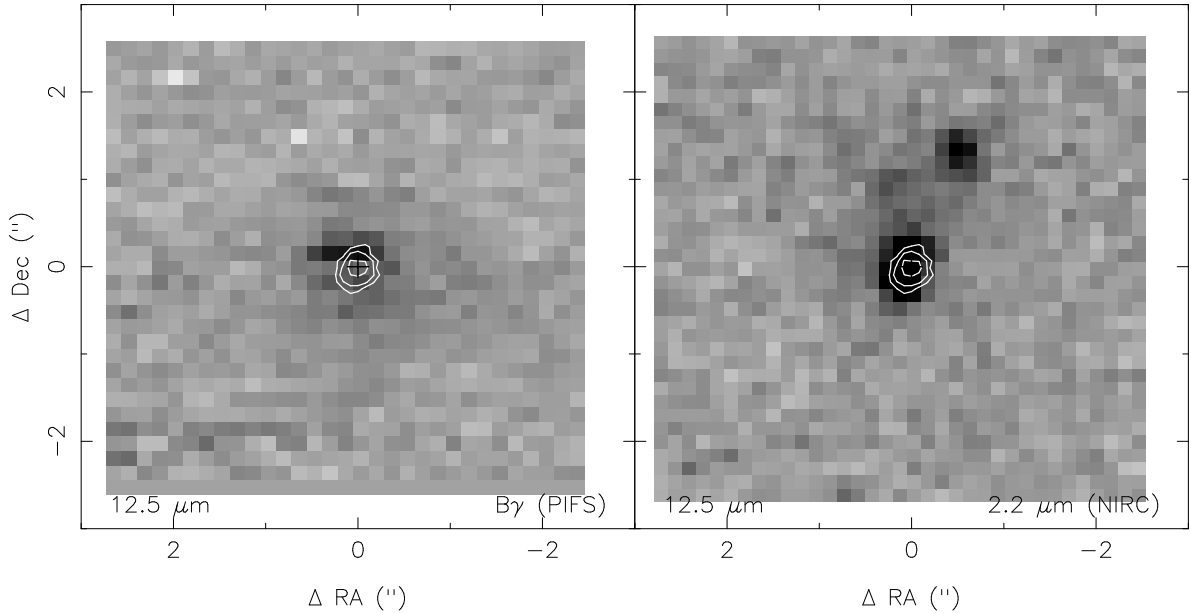}}
\caption[]
%{\  The Keck 12.5\m\ contours enclosing regions where the emission is greater than 90\%, 67\%, and 50\% of the peak flux overlaid on a) the PIFS Br$\gamma$ data, b) the PIFS 2.2\m\ continuum data, and c) the Keck NIRC 2.2\m\ continuum data.  To $\sim$0\farcs5 accuracy, the peak of the 12.5\m\ data is aligned with the peak of the Br$\gamma$ and 2.2\m\ continuum emission (see the associated discussion in the text).  All frames are oriented with north up and east to the left.}
{\  The Keck 12.5\m\ contours enclosing regions where the emission is greater than 90\%, 67\%, and 50\% of the peak flux overlaid on the PIFS Br$\gamma$ data and the Keck NIRC 2.2\m\ continuum data.  The 2.2\m\ emission contains multiple unresolved components that presumably correspond to the super star clusters described in Thuan et al. (1997).  Table~\ref{tab:size} provides a comparison of the size of \sbs\ at different wavelengths.  To $\sim$0\farcs5 accuracy, the peak of the 12.5\m\ data is aligned with the brightest peak of the 2.2\m\ continuum and Br$\gamma$ emission (see the associated discussion in the text).  Both frames are oriented with north up and east to the left and the central coordinates are those given in Table~1.}
\label{fig:overlays}
\end{figure}

The PIFS observations presented in Figure~\ref{fig:sbs0335_brgamma} show that the Br$\gamma$ emission is apparently spatially extended: the line emission spans 1\arcsec\ FWHM (260~pc) compared with the 0\farcs5-0\farcs8 seeing at that wavelength.  Whereas the signal-to-noise ratio of the PIFS continuum data is too low for us to accurately gauge by eye the center of the continuum distribution, a central moment analysis of the continuum and line emission maps yields a negligible ($<$0\farcs03; $<10$~pc) continuum-line offset; there is no evidence in the integral field spectroscopy data for the 60~pc offset between the Br$\gamma$ and 2\m\ continuum peaks claimed by Vanzi et al. (2000).  By the design of the PIFS (Murphy, Matthews \& Soifer 1999), the Br$\gamma$ and 2.2\m\ continuum images are spatially tied together.  If the ad hoc assumption is made that the 12.5\m\ emission coincides with the Br$\gamma$ emission, this ties the 12.5\m\ emission and the 2.2\m\ continuum emission of \sbs\ together.  Figure~\ref{fig:overlays} illustrates this assumption, and thus orients the 12.5\m\ emission of \sbs\ relative to that at Br$\gamma$ and 2.2\m.  Both this and the offsetting measurements previously described are consistent with the 12.5\m\ emission of \sbs\ coinciding with the bright peak in the 2.2\m\ continuum.  Using 0\farcs33 resolution $Ks$ imaging, Vanzi et al. (2000) note a $\sim$1\farcs5 separation between two distinct components along the southeast-northwest direction, and $HST$ WFPC2 $V$ and $I$ imaging show several distinct clumps spanning $\sim$4\arcsec\ along a southeast-northwest orientation (Papaderos et al. 1998).  Such substructure is seen in the Keck NIRC image, but the PIFS near-infrared data give only a faint hint of the multiple peaks, presumably due to the limited spatial resolution and sensitivity of the integral field spectroscopy data.
%actual difference in Brgamma and 2mu continuum is 0.162 pixels=0.027"=7pc

No velocity structure is resolved for the Br$\gamma$ emission.  This is not surprising since the resolution is of order 200\kms\ and the VLA \HI\ data of Pustilnik et al. (2001) only show $\sim$40\kms\ solid-body rotation for the eastern \HI\ peak that is centered on \sbs.  Furthermore, the lack of significant velocity structure of the Br$\gamma$ line is consistent with the small velocity dispersion ($<10$\kms) of the optical emission lines (Izotov et al. 1997).
%mention that large-scale winds in starburst galaxies can exceed 300-1000\kms?
%Heckman, T.M., Armus, L. \& Miley, G.K. 1987, \aj, 92, 276

Measures of the starburst timescale can be inferred from near-infrared continuum data.  However, Vanzi et al. (2000) find that the combined helium, H$_2$, and Br$\gamma$ emission lines contribute about 50\% of the $K$-band flux (within their 1\arcsec$\times$0\farcs8 spectroscopy slit), and that Br$\gamma$ alone accounts for 6-7\%.  To within 10\% certainty, the integrated Br$\gamma$ flux from the PIFS data is $6 \times 10^{-18}$~W~m$^{-2}$, or approximately 4\% of the global $K$-band flux.   The Br$\gamma$ equivalent width is also indicative of the starburst age (\S~\ref{sec:sf_age}) and is not contaminated by other nebular emission lines.  From the PIFS data the Br$\gamma$ equivalent width is 235$\pm$33\AA.  A cursory review of the literature did not find a larger extragalactic Br$\gamma$ equivalent width.
%mK=15.43; f(K)=0.42 mJy or 2.786e-16 W/m^2/micron
%delta lambda f_lambda=1.453e-16 W/m^2

\subsection{Radio Continuum Results}
\label{sec:radio_results}
\begin{figure}[!ht]
\centerline{\psfig{figure=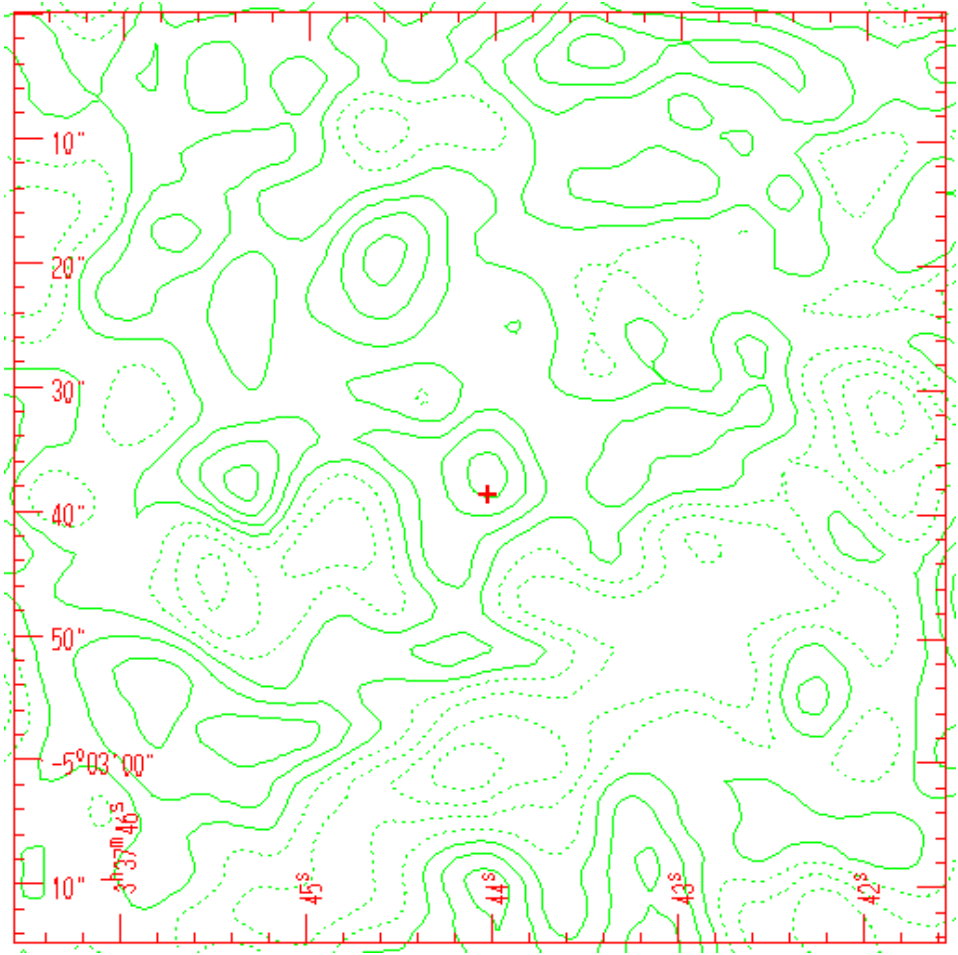,width=5in,bbllx=51pt,bblly=137pt,bburx=559pt,bbury=645pt}}
\caption[]
{\ 1.49 GHz data from the VLA archive.  The 3$\sigma$ source located at the center of this field is at the nominal optical position of \sbs, which is indicated by a cross.  The contour levels are plotted at 0.15~mJy~beam$^{-1}$ intervals, starting at 0.45~mJy~beam$^{-1}$.  Coordinates are given in the epoch J2000.0.}  
\label{fig:radio}
\end{figure}

\sbs\ is unresolved in the high resolution map (FWHM=6\arcsec\ beam) at 1.49~GHz (Figure~\ref{fig:radio}).  There is a 3$\sigma$ source with a peak brightness of $0.40~{\rm mJy}~{\rm beam}^{-1}$ at the nominal optical position of \sbs; the radio source has a flux density of 0.40~mJy.  Each coordinate has an rms error of about 1\arcsec\ due to noise, but this is much better than the NVSS positional uncertainty of $\sim$7\arcsec\ for faint sources like \sbs.  No extended radio continuum emission is visible, but a $2.3\pm0.4$~mJy source (the NVSS total flux) smooth source 30\arcsec\ in size would have a brightness of only $0.1~{\rm mJy}~{\rm beam}^{-1}$ and be completely undetectable.  Indeed, no 20 cm continuum emission is detected in a VLA survey with 1~mJy~beam$^{-1}$ (4$\sigma$) sensitivity and a 20\farcs5$\times$14\arcsec\ beam (Pustilnik et al. 2001).  On the other hand, the offset of the NVSS source from the optical position (12\arcsec\ at a position angle of 50\arcdeg\ east of north) is large enough that the NVSS source may well be a blend of \sbs\ and two other faint sources to the northeast visible in the contour plot.  In short, the flux of the nearby NVSS source probably overestimates the total flux of \sbs, but the 0.40~mJy component probably corresponds to the compact part of \sbs, particularly since they agree well in position.  The chance that this is a non-detection and is instead a random 3$\sigma$ noise bump coincidentally close to the nominal position of \sbs\ is low.  The probability of finding an unrelated radio source with flux $\geq0.4$~mJy within 6\arcsec\ is less than $2\times10^{-3}$ (Langston et al. 1990).
%their equation 2

\subsection{CO Results}

\begin{figure}[!ht]
\centerline{\psfig{figure=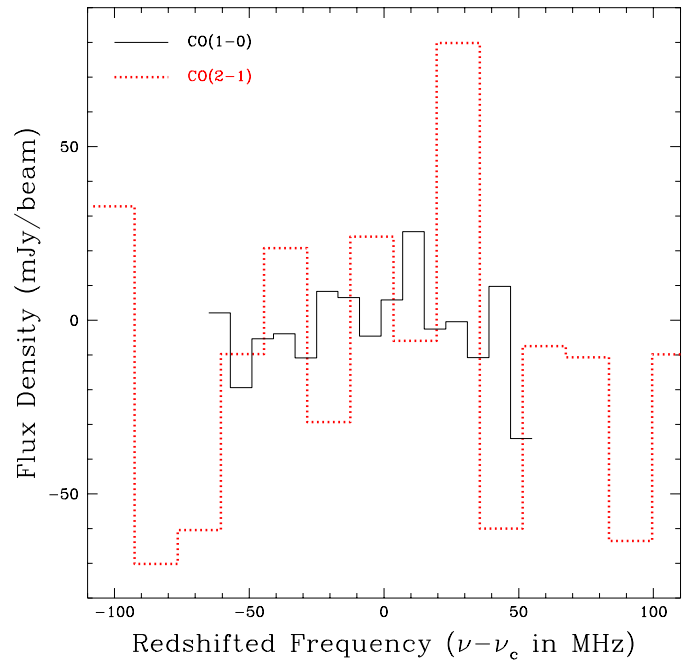,width=3.5in,bbllx=18pt,bblly=153pt,bburx=563pt,bbury=689pt}}
\caption[]
{\ The CO(1-0) and CO(2-1) spectra, centered at $\nu_{\rm c}=113.73$~GHz and $\nu_{\rm c}=227.44$~GHz, respectively.  The observations assumed an optical redshift of 4076\kms\ in the Local Group reference frame (Thuan, Izotov \& Lipovetsky 1997).  The rms level for the (1-0) data is 12~mJy, and 30~mJy for the (2-1) line.}
\label{fig:CO_spec}
\end{figure}

%As seen in Figures~\ref{fig:CO_spec} and \ref{fig:CO_map}, neither CO nor millimeter continuum emission was detected.  The rms limits of 12~mJy and 30~mJy were achieved for the CO(1-0) and CO(2-1) observations, respectively, from an average over four spectral channels (Figure~\ref{fig:CO_spec}).  Figure~\ref{fig:CO_map} shows the integrated CO maps averaged over the central 100\kms\ covering the \HI\ line width (Thuan, Izotov \& Lipovetsky 1997).  No significant emission 
As seen in Figure~\ref{fig:CO_spec}, neither CO nor millimeter continuum emission was detected.  The rms limits of 12~mJy and 30~mJy were achieved for the CO(1-0) and CO(2-1) observations, respectively, from an average over four spectral channels (Figure~\ref{fig:CO_spec}).  No significant emission or absorption features were detected anywhere in the field.  Since the synthesized beams are larger than the bright central starburst regions, it is not expected that the interferometer has resolved out the CO emission.  Given that the CO line widths are typically only half the \HI\ line widths in metal-poor galaxies (Sage et al. 1992), a CO line width of approximately 50\kms\ would be expected for \sbs.  With this adopted line width, the $3\sigma$ upper limits are $S_{\rm CO(1-0)} < 1.2$~Jy\kms\ and $S_{\rm CO(2-1)} < 3.0$~Jy\kms\ for the integrated CO line fluxes.  For the continuum data, the 1~GHz bandwidths in both the upper and lower side bands were averaged, resulting in $3\sigma$ upper limits of $f_\nu(112~{\rm GHz})< 2.3$~mJy and $f_\nu(229~{\rm GHz})< 11$~mJy.

Given the observational limits, the CO(1-0) data provide a more stringent constraint on the molecular hydrogen gas mass than the CO(2-1) data---blue compact galaxies typically have brightness temperature ratios of $T_b(2-1)/T_b(1-0)=0.3-0.6$ (Sage et al. 1992).  We thus concentrate on the CO(1-0) data for the remainder of this section.  The integrated CO(1-0) flux in Jy\kms\ is related to the molecular hydrogen gas mass by $M({\rm H}_2)=1.2\times10^{4}S_{\rm CO(1-0)}(d/{\rm Mpc})^2 (\alpha/\alpha_{\rm G})$\msun, where $\alpha_{\rm G}$ is the Galactic CO to H$_2$ conversion factor (Wilson 1995).  Adopting the Galactic conversion factor results in $M({\rm H}_2) < 4.1\times 10^{7}~M_{\odot}~(3\sigma)$.  However, the Galactic conversion factor is thought to significantly underestimate the molecular hydrogen gas mass in the metal-poor galaxies.  Estimating the proper value of $\alpha$ for \sbs\ is somewhat problematic since CO has yet to be detected in such a metal-deficient system.  For galaxies with metallicities between solar and 0.1 solar, Wilson (1995) finds $\alpha \propto Z^{-0.67}$.  Extrapolating this relationship to a metallicity of 1/41 solar yields $\alpha \simeq 12 \alpha_{\rm G}$ and $M({\rm H}_2) < 4.9\times 10^{8}$\msun\ for \sbs.  However, these estimates are only applicable for dense regions where the CO molecules could survive the intense ultraviolet radiation field.  In regions of low extinction ($A_V \lesssim2$~mag), the global CO/H$_2$ abundance ratio is expected to be extremely non-linear with metallicity (Maloney \& Black 1988).

\section{Discussion}

\subsection{Dust and The Infrared Spectral Energy Distribution}
\label{sec:sed}
Thuan, Sauvage \& Madden (1999) fit the observed mid-infrared spectrum with the profile for a 250~K extincted blackbody with emissivity $Q_\nu \propto \nu^{\beta}$ where $\beta=1.5$ is the assumed dust emissivity index.  Data at longer wavelengths can help better constrain the infrared spectral energy distribution: in addition to the new flux measurement at 67\m, there are upper limits available from \IRAS\ at 12, 25, 60, and 100\m. These limits are compatible with the levels detected at similar wavelengths (see Figure~\ref{fig:fit}).  With the infrared picture of \sbs\ now more fully developed, it is clear that the data suggest multiple dust components.  A fit of the superposition of two blackbodies with emissivity $Q_\nu \propto \nu^{1.5}$, with variable normalizations and dust temperatures, to the 6.7, 7.7, 12.5, 25, and 67\m\ broadband data results in dust temperatures for the two components of $T_{\rm d}=80$~K and 210~K.\footnote{\IRAS\ marginally detected (2$\sigma$) \sbs\ at 25\m.  If a 2$\sigma$ 25\m\ flux is used in the fit instead of a 3$\sigma$ upper limit, the resulting dust temperatures are 70~K and 207~K.}  Varying the emissisivity from $Q_\nu \propto \nu^{1.0}$ to $\nu^{2.0}$ changes the dust temperatures by a maximum of $\pm$20\%.  Two relatively warm dust components are also evident in Seyfert galaxies (P\'{e}rez Garc\'{\i}a, Rodr\'{\i}guez Espinosa \& Santolaya Rey 1998) and luminous galaxies such as Arp 220, NGC 6240 and NGC 4038/39 (Klaas et al. 1997).  More far-infrared data are needed to verify whether \sbs\ has an additional colder ($T_{\rm d}\sim25$~K) dust component.

The density and total amount of dusty material in the core of \sbs\ determines the central opacity.  A simple estimate of the optical depth is the ratio of the observed surface brightness to the intensity of the expected blackbody profile, $\tau \sim f_\nu \Omega^{-1}/B_\nu(T_{\rm d})$, where $\Omega$ is the solid angle of the object.  The mid-infrared opacity for \sbs\ is less than $3 \times 10^{-3}$ at 12.5\m\ for a wide range of hot dust temperatures ($T_{\rm d}^{\rm hot}>110$~K).  Such an opacity corresponds to $A_V<1$~mag, in agreement with the result of Izotov et al. (1997).  Furthermore, the total ionizing luminosity computed from the number of UV photons per second (\S \ref{sec:sf_age}) is of order $8 \times 10^8~L_\odot$, similar to the mid-infrared luminosity (Table 1).  These results suggest $\tau_{\rm uv}\sim1$, in contrast to the claim for $A_V\sim20$~mag by Thuan, Sauvage \& Madden (1999), and argue that the model of a hot blackbody affected by a large column of extinction is invalid.
%1.5e53 photons/sec * hc/100nm = 7.8e8 Lsun

Since the infrared radiation is optically thin in emission, dust masses can be directly computed using the infrared fluxes.  The infrared flux density may be expressed in terms of the number of grains $N$ of size $a$, density $\rho$, and emissivity $Q_\nu$ radiating at temperature $T_{\rm d}$ at distance $d$:
$
f_\nu = N \pi B_\nu(T_{\rm d}) Q_\nu 4 \pi a^2 / 4 \pi d^2.
$
And since the dust mass can be expressed as $M_{\rm dust}=N {4 \over 3} \pi a^3 \rho$, one finds
\be
M_{\rm dust}={4 a \rho f_\nu d^2 \over 3 Q_\nu B_\nu(T_{\rm d})} 
\equiv {f_\nu d^2 \over \kappa_\nu B_\nu(T_{\rm d})}
\ee
(e.g. Hildebrand et al. 1977).  In other words, the dust mass is a function of the observed flux density and the estimated dust temperature, and does not depend on the size of the emitting region.  If the dust temperature is 210~K for the hot dust component and the opacity coefficient $\kappa(12$\m)$=940~{\rm cm}^2~{\rm g}^{-1}$ (Li \& Draine 2001), the dust mass computed from the 12.5\m\ flux is $M^{\rm hot}_{\rm dust}\simeq6~M_\odot$.  The mass of the cooler dust may be computed from the dust emission at 67\m, radiation which is more likely to arise from steady state processes compared to the mid-infrared emission from very small grains and polycyclic aromatic hydrocarbons (PAHs).  The total mass of the presumably larger and cooler dust grains in \sbs, if $T_{\rm d}=80$~K and $\kappa(70$\m)$=67.3~{\rm cm}^2~{\rm g}^{-1}$ (Li \& Draine), is $M^{\rm warm}_{\rm dust}\simeq2400~M_\odot$.  Thus the warm dust mass is much larger than the hot dust mass in a simple two-component model.  We will henceforth use it to represent the total dust mass in \sbs.  Considering that the long wavelength end of the spectral energy distribution is poorly constrained, the warm dust temperature estimate, and thus the overall dust mass, is highly uncertain.  Any cold dust, which is unconstrained by the measurements reported here, would likely dominate the total dust mass. 
%2400 to be exact

\ISO\ observations show a mid-infrared spectrum with none of the standard PAH features (Thuan, Sauvage \& Madden 1999).  For a two-component dust model, the hotter dust in \sbs\ is 400 times less massive than the warm dust.  Unless it can be explained by sensitivity arguments, it seems peculiar then that there are no obvious PAH features in its mid-infrared spectrum---typically only regions with very energetic interstellar radiation fields, through destruction and/or ionization processes, lack the ubiquitous mid-infrared aromatic emission features (e.g. the core of M~17; Cesarsky et al. 1996).  Conversely, emission from merely warm dust should be replete with aromatic features.   A possible explanation is that PAHs are thought to condense out of the outflows of carbon-rich red giants (Allamandola, Tielens \& Barker 1989), but the starburst in \sbs\ is too young ($<7$~Myr) for stars to have reached this phase---only stars more massive than $\sim20~M_\odot$ are expected to spend less than 10~Myr on the main sequence (e.g. Bressan et al. 1993).

\subsection{The Gas-to-Dust Ratio}
\label{sec:gas_to_dust}

The global gas-to-warm dust mass ratio is relatively straightforward to compute.  From the \HI\ measurement of Pustilnik et al. (2001) of $M({\hbox{H \sc i}}) \sim 8 \times 10^8~M_\odot$ and the molecular hydrogen mass limit of $M({\rm H}_2) < 4.9\times 10^{8}$\msun,
\be
\left.{M_{\rm gas} \over M_{\rm dust}}\right|_{\rm global} = \left.{M({\hbox{H \sc i}})+M({\rm H_2}) \over M_{\rm dust}}\right|_{\rm global} \lesssim 10^{5.7}.
\ee
%(8e8+4.1e7)/2400=3.50e5  (8e8+4.9e8)/2400=5.38e5
This value lies on the high end of the expected range for a compact dwarf galaxy compared with the range $10^3 \lesssim {M_{\rm gas} \over M_{\rm dust}} \lesssim 10^5$ typically seen for blue compact dwarfs (Lisenfeld \& Ferrara 1998), and to that usually found in more mundane regions, like the disks of spiral galaxies where ${M_{\rm gas} \over M_{\rm dust}} \sim 300$ (see Mayya \& Rengarajan 1997 and references therein).

However, a global gas-to-dust ratio for \sbs\ is a misleading quantity since the dust and gas are not measured in the same volume.  A more sensible constraint on the gas-to-dust mass ratio is derived from the ionized gas mass.  The ionized gas density within the Str\"{o}mgren sphere is $n_{\rm e} \sim 45~{\rm cm}^{-3}$ (\S \ref{sec:NIR_data}).  The central 80~pc therefore contains approximately $3 \times 10^5~M_\odot$ of gas, and
\be
\left.{M_{\rm gas}^{\rm ionized} \over M_{\rm dust}}\right|_{r<80 {\rm pc}} \gtrsim 100,
\ee
representing a lower limit on the central gas-to-dust mass ratio.  This estimate is limited by the assumption that the entire $\sim2400~M_\odot$ dust mass falls within the inner 80~pc.

\subsection{Clues from the Br$\gamma$ Emission}
\label{sec:sf_age}
The near-infrared broadband fluxes of starbursts can be dominated by nebular emission lines that unduly influence stellar age estimates (Leitherer \& Heckman 1995).  An alternative indicator of the starburst age is the Br$\gamma$ equivalent width, a parameter that reflects the ratio of young blue massive stars to the evolved red stellar population.  For a plausible range of initial mass function parameters, the Starburst99 models (Leitherer et al. 1999) indicate that $EW$(Br$\gamma)\simeq235 \AA$ for $Z_\odot/50$ metallicity material corresponds to a stellar age of $3-6 \times 10^6$ years.  This age holds for both an instantaneous and a continuing star formation model, consistent with the age estimate from other near-infrared spectroscopic and broadband data (Vanzi et al. 2000).  

If the central \HII\ region is modeled as a Str\"{o}mgren sphere, it is possible to use the size of the Br$\gamma$ emission and the Br$\gamma$ luminosity to infer the density of the ionized gas $n_{\rm e}$.    For an intrinsic Br$\alpha$-to-Br$\gamma$ luminosity ratio of 2.9, the Br$\gamma$ luminosity for \sbs\ corresponds to $N_{\rm uv} = {N_{\rm uv} \over N_{\rm Br\alpha}} {N_{\rm Br\alpha} \over N_{\rm Br\gamma}} N_{\rm Br\gamma} = {N_{\rm uv} \over N_{\rm Br\alpha}} {L_{\rm Br\alpha} \lambda_{\rm Br\alpha} \over L_{\rm Br\gamma} \lambda_{\rm Br\gamma}} {L_{\rm Br\gamma} \over h\nu_{\rm Br\gamma}} \sim 1.5 \times 10^{53}$ ionizing photons per second (see Ho, Beck \& Turner 1990), or approximately 2800 O5 stars (e.g. Panagia 1973).  Since the volume of the Str\"{o}mgren sphere can be expressed as ${4 \over 3}\pi R^3_{\rm Br\gamma} = {N_{\rm uv} / \alpha_{\rm B} n_{\rm e}^2}$, for a measured $R_{\rm Br\gamma}=130$~pc the inferred density is $n_{\rm e} \sim 45~{\rm cm}^{-3}$, reasonable for \HII\ regions (Osterbrock 1989).  Here we have assumed Case B recombination (with $\alpha_{\rm B} \simeq 2.6 \times 10^{-13}~{\rm cm}^{-3}~{\rm s}^{-1}$) for a nebula at temperature $T_{\rm e} = 10,000$~K, but the same electron density is found to within a factor of two for a wide temperature range.

Inasmuch as Br$\gamma$ emission is associated with \HII\ regions, it is indicative of the heating input to the interstellar medium; the Br$\gamma$ flux can be related to the thermal radio flux density $f_\nu^{\rm th}$.  Following Ho et al. (1989) and assuming a radio spectral slope of $-0.1$ appropriate for optically thin free-free emission from ionized gas, we estimate the expected thermal radio continuum flux at 20~cm from $f_\nu^{\rm th}(20~{\rm cm}) \approx 1.0~{\rm mJy} \left({d \over {\rm Mpc}}\right)^{-2} \left({N_{\rm uv} \over 10^{50}~{\rm s}^{-1}}\right)$.  The number of ionizing photons per second inferred from the integrated Br$\gamma$ flux for \sbs\ predicts $f_\nu^{\rm th}(20~{\rm cm})\approx 0.5$~mJy, approximately the observed value for $f_\nu(20~{\rm cm})$.  Since the prediction is at least as large as the observed 20~cm flux density, the straightforward interpretation is that there is little or no evidence for nonthermal radio contributions---there has been, at most, a low level of supernova activity in the short star-formation history of \sbs.  A quick 5~GHz C-array observation with the VLA could easily verify whether the radio source has a flat spectrum indicative of \HII\ regions or a steep spectrum produced by supernova remnants.  For comparison, the global thermal fraction of the flux density at 20~cm in normal star-forming galaxies is about 10\% (Condon 1992).  According to Matteucci \& Recchi (2001), the peak in the supernovae rate (and thus the maximum metal enrichment rate) for an instantaneous starburst episode does not occur until 40-50~Myr after the onset of star formation.  Such an upper limit to the age of the starburst in \sbs\ agrees with the constraints placed by the Br$\gamma$ data.  Furthermore, a scenario of modest prior supernova activity would also be consistent with the interpretation laid out in Section~\ref{sec:ir_to_radio} for the high infrared-to-radio ratio for \sbs.

\subsection{The Infrared-to-Radio Ratio}
\label{sec:ir_to_radio}

\begin{figure}[!ht]
\centerline{\psfig{figure=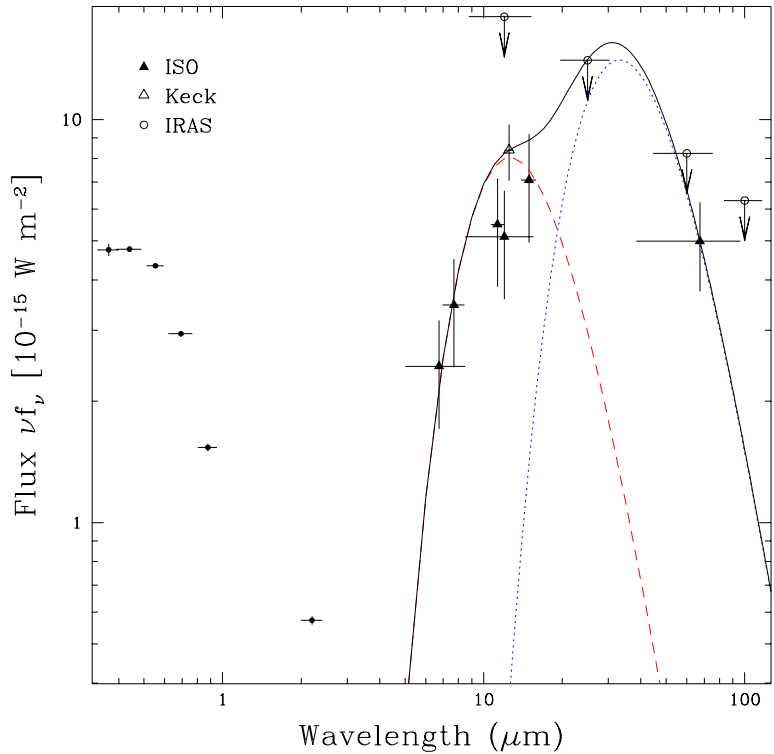,width=4in,bbllx=20pt,bblly=155pt,bburx=564pt,bbury=689pt}}
\caption[]
{\ A fit to the infrared spectral energy distribution using a superposition of two $Q_\nu \propto \nu^{1.5}$ blackbodies, one at $T_{\rm d}=210$~K (dashed line) and the other at $T_{\rm d}=80$~K (dotted line).  The upper limits are from \IRAS, and are equal to three times the rms deviation of 7-10 scans from the median scan.  The five mid-infrared broadband values from Thuan, Sauvage \& Madden (1999) have been assigned 30\% flux uncertainties since (i) that is the discrepancy between their spectroscopy and broadband photometry and (ii) 30\% is also the approximate calibration uncertainty for \ISO\ mid-infrared data.}  
\label{fig:fit}
\end{figure}

The radio continuum flux for \sbs\ spatially coincides with its emission at 12.5\m.  Thus, if the emission in the mid- and far-infrared is spatially coincident, the far-infrared-to-radio ratio is $f_\nu(67~\micron)/f_\nu(1.49~{\rm GHz}) \simeq 280$.  This flux density ratio is compatible with the typical ratio for normal star-forming galaxies, which is approximately $200^{+120}_{~-70}$.  Note, however, that in the absence of \IRAS\ detections for \sbs\ an \ISO-based $f_\nu(67~\micron)$ has been used as a proxy for the typical \IRAS-based definition for the far-infrared ``FIR'' flux (Helou et al. 1988), which effectively recovers the entire 42-122\m\ far-infrared flux; the ISOPHOT filter used here, on the other hand, has a shorter central wavelength and an overall narrower profile (\S \ref{sec:ISO}) and thus does not provide as robust an estimate of the bolometric infrared flux.  The FIR 42-122\m\ flux for \sbs\ computed from the model spectral energy distribution shown in Figure~\ref{fig:fit} is $5.4\pm1.1 \times 10^{-15}~{\rm W}~{\rm m}^{-2}$, and thus the logarithmic measure defined by Helou, Soifer \& Rowan-Robinson (1985) is $q = \log\left[ {{\rm FIR} \over 3.75 \times 10^{12}~{\rm W~m}^{-2}} \right] - \log \left[ {f_\nu(1.49~{\rm GHz}) \over {\rm W~m}^{-2}~{\rm Hz}^{-1}} \right] \simeq 2.56\pm0.09$.  The 20\% uncertainty is conservatively derived from the full range of FIR model fluxes found after invoking different combinations of the 25\m\ flux upper limit (2$\sigma$ to 3$\sigma$) and the dust emissivity index ($1.0\lesssim\beta\lesssim2.0$).  In other words, the far-infrared-to-radio ratio estimated from the model spectral energy distribution is $\simeq360\pm70$, slightly outside the $1\sigma$ range for normal star-forming galaxies.  

Given its youth, small size, and the inferred lack of nonthermal radio flux, it is surprising that \sbs\ exhibits a fairly `normal' infrared-to-radio ratio.  In resolved sources, the radio continuum emission is usually more diffuse than the infrared light (Marsh \& Helou 1998), and is explained by the spreading of cosmic rays as they decay and are trapped in magnetic field lines (Helou \& Bicay 1993).  The smearing scalelength in normal galaxies is many hundreds of parsecs (Marsh \& Helou 1998), much larger than the 80~pc mid-infrared size of \sbs.  Moreover, the relatively small disk scale heights for dwarf galaxies may result in cosmic rays escaping comparatively easily (Helou \& Bicay 1993).  Finally, the youth of \sbs\ means that relatively few supernovae have occurred, resulting in low cosmic ray production and weak interstellar magnetic fields.  All these lines of reasoning predict a low nonthermal radio continuum flux for \sbs\ compared to the far-infrared flux, consistent with the findings using the Br$\gamma$ data.  In short, since the canonical far-infrared-to-radio ratio presumably hinges on a typically strong ($\sim90$\%) nonthermal contribution to the radio flux, the apparently negligible nonthermal radio emission of \sbs\ should result in an extremely high far-infrared-to-radio ratio. 

%The infrared-to-radio ratio is even more difficult to interpret if the coarser NVSS map (FWHM=45\arcsec\ beam) is used, which picks up almost six times as much flux as the high resolution map; the far-infrared-to-radio ratio is closer to 40 if the entire 1.49~GHz flux from the NVSS mapping is used.  In other words, based simply on expectations for the far-infrared to radio ratio, the larger radio continuum flux in the NVSS map is consistent with contributions from two sources external to the inner 80~pc (see Figure~\ref{fig:radio}), plus diffuse emission.
%Condon quotes 7" RA,DEC uncertainty for S=2.3mJy, and 0.5" for S>30mJy

Given the elevated mid- to far-infrared ratio and the unusual infrared spectral energy distribution of \sbs, perhaps it is more appropriate to instead compute a total infrared-to-radio ratio.  In normal galaxies the far-infrared flux is a useful measure of the total infrared flux (TIR), with FIR typically half of the 3-1100\m\ bolometric infrared flux (Dale et al. 2001): TIR/radio $\sim400^{+230}_{-150}$.  But the conspicuously intense interstellar radiation field of \sbs\ results in an unusually strong contribution at mid-infrared wavelengths---the FIR flux for \sbs\ computed from the model spectral energy distribution shown in Figure~\ref{fig:fit} (and all plausible variations) is less than one fourth the model's bolometric infrared flux, resulting in TIR/radio $\sim1600\pm300$.  In short, the total infrared to radio flux for \sbs\ is about four times that in normal star-forming galaxies, and is likely due to a relative lack of trapped interstellar cosmic rays, resulting in comparatively weak radio emission.

\subsection{The Intensity of the Interstellar Radiation Field}

In a detailed study of the N~4 complex of \HII\ regions in the Large Magellanic Cloud, Contursi et al. (1998) show how the mid-infrared color $f_\nu(15~\micron)/f_\nu(7~\micron)$ strongly varies as a function of interstellar radiation field intensity, with the highest ratios ($f_\nu(15~\micron)/f_\nu(7~\micron) \gtrsim 4$) corresponding to ultraviolet radiation densities approaching $10^4$ times the interstellar radiation field in the solar neighborhood.  Though the distance to \sbs\ precludes such a spatially-detailed analysis, the average interstellar heating intensity can be estimated from $\left<U\right> \sim L(12.5~\micron)/4 \pi r^2 \approx 15~{\rm erg}~{\rm cm}^{-2}~{\rm s}^{-1} \simeq 10,000~G_o$ where $G_o=1.6 \times 10^{-3}~{\rm erg}~{\rm cm}^{-2}~{\rm s}^{-1}$ and $1.7~G_o$ is the flux of the local interstellar far-ultraviolet radiation field (Hollenbach \& Tielens 1999 and references therein).  This result is consistent with the findings of Contursi et al. (1998; 2000): an unusually high mid-infrared color, $f_\nu(15~\micron)/f_\nu(7~\micron)~\approx~6.4$ in the case of \sbs\ (Thuan, Sauvage \& Madden 1999), corresponds to an interstellar radiation field at least four orders of magnitude greater than the local value.  Moreover, a reasonable dust temperature is found for \sbs\ if it is assumed that $\left<U\right> \propto T_{\rm d}^{4+\beta}$.  Scaling from the $\sim$17.5~K dust temperature found for diffuse Galactic cirrus (Boulanger et al. 1996), an order of magnitude estimate for the interstellar dust temperature for such a radiation field is $T_{\rm d} \sim (17.5$~K$) (10^4~G_o/G_o)^{1 \over 4+\beta} \sim 100$~K.  Though this result is compatible with the dust temperature predicted for the bulk of the interstellar medium in \sbs\ (Section~\ref{sec:sed}), it is the hot dust that is primarily emitting at 12.5\m.  One interpretation of this discrepancy is that the small grain emission at mid-infrared wavelengths results from a semi-stochastic process; the small grain emission is perhaps not well-represented by a hot blackbody (e.g. Draine \& Anderson 1985).
%if use 67.3 micron flux (which is problematic since we don't know the size at that wavelength....) we get 8.85 erg/cm^2/s -> 10^3.75 Go -> 76.25 Kelvin dust.

\subsection{The Compact Nature of the Starburst Region}

The mid-infrared data presented here, with angular resolution better than 0\farcs4, represent the highest spatial resolution achieved to date in probing the blue compact dwarf galaxy \sbs.  \sbs\ is compact at 12.5\m\ compared with its stellar and gaseous constituents.  In normal star-forming galaxies the extent of the mid-infrared, far-infrared, and submillimeter emission is comparable to the optical size (Dale et al. 2000; Rice et al. 1988; Chini et al. 1995), and detailed modelling of the optical and near-infrared emission from normal late-type spiral galaxies shows that dust scalelengths typically are larger than that of the stars and that dust radially extends to larger distances than stars in the disk (Xilouris et al. 1999).  On the other hand, the high angular resolution studies of Soifer et al. (2000) and Soifer et al. (2001) show that it is common for both ultraluminous and starburst galaxies to be particularly compact at 12\m\ compared with their optical distributions, especially for ULIRGs, with typically more than 50\% of the mid-infrared emission arising from $<100-300$~pc-sized regions.  Thus perhaps it is relatively common for actively star-forming systems to produce the bulk of their thermal mid-infrared emission in small volumes surrounding a handful of bright super star clusters.  This may be especially true for dwarf systems like \sbs, where a relatively small stellar population may be dominated by a few extremely bright stars or star clusters.

%The compact nature of the mid-infrared emission from \sbs\ is similar to that found in much larger systems, but it still does not explain why the ionized gas is spread over a much larger region than the $\sim$80~pc distribution of hot dust.  Can the discord be explained through diffusion timescale arguments?  In other words, what is the dispersive speed for the interstellar dust if most of the mid-infrared emission has come from a single region?  For example, the brightest two of the six most luminous star clusters described by Thuan, Izotov \& Lipovetsky (1997) are only 80~pc apart.  Assuming a constant dispersion, the necessary expansion speed is $40$~pc/5~Myr $\sim8$\kms, similar to the velocity dispersion of the ionized gas in \sbs\ (Izotov et al. 1997) and roughly what is expected for the dispersal speed of the material in giant molecular clouds, objects which span size scales similar to the mid-infrared emission of \sbs\ (D. Hollenbach, private communication).  However, given the evidence from $HST$ imaging for at least one supershell spanning hundreds of parsecs (Thuan, Izotov \& Lipovetsky 1997), it seems unlikely that dust has only been spread through simple diffusion.  Indeed, Thuan, Izotov \& Lipovetsky (1997) find signatures of dust reddening throughout the $\sim520$~pc star-forming region.  

The lack of extended hot dust emission in \sbs\ is most easily attributed to geometry:  the relatively small number of bright stars in \sbs\ heat to high temperatures only the most proximate dust.  As outlined in Section~\ref{sec:sed}, the mass of this hot dust is $\sim 6~M_\odot$.  The bulk of the $\sim 2400~M_\odot$ of dust spread throughout the rest of the interstellar medium is located significantly farther from the hottest stars, and is thus only warmed to $\sim80$~K on average.  Balancing the radiation absorbed and reemitted by dust grains yields the following simple relation for the approximate temperature of dust grains in thermal equilibrium with the surrounding interstellar radiation field:
\be
T_{\rm d}=109~{\rm K} ~ \left({Q_{\rm uv} \over Q_{\rm ir}}\right)^{\onequarter} \left({\rm d} \over {\rm 1~pc}\right)^{-\onehalf}  \left(L_{\rm uv} \over 10^9~L_\odot\right)^{\onequarter}
\ee
where $Q_{\rm uv}$ and $Q_{\rm ir}$ are the absorption and emission coefficients in the ultraviolet and infrared, respectively.  If the dust emission is optically thin and distributed in a circularly symmetric fashion centered on the $10^9~L_\odot$ ionizing luminosity source (e.g. the brightest super star cluster), then the hot dust is $\sim3$~pc from the ionizing source and the warm dust lies at a characteristic distance of $\sim15$~pc (assuming $Q_\lambda \propto \lambda^{-1.5}$).  A substantial portion of the ionizing photons from the star clusters, on the other hand, encounter neutral gas at much larger distances, and that is why the central \HII\ region in \sbs\ spans more than 1~kiloparsec (Izotov et al. 1997).  Needless to say, the angular resolution of the extant infrared data is at least an order of magnitude too low to corroborate this scenario. 

In many respects, \sbs\ resembles NGC~5253: it is a low metallicity dwarf galaxy dominated by a $<$10~Myr old compact and strong mid-infrared emitting star-forming region for which the radio continuum emission is almost entirely thermal (Beck et al. 1996; Gorjian, Turner \& Beck 2001).  Can a useful comparison be made between the star-forming regions in NGC~5253 and \sbs?  Probably not, in that the mid-infrared size of \sbs\ is truly extended and is not confined to an ultracompact 1-2~pc super star cluster containing several thousand O stars as in NGC~5253 (Turner et al. 2000)---the stellar component of \sbs\ is comprised of multiple star clusters spread throughout the inner 520~pc (Thuan, Izotov \& Lipovetsky 1997).  Moreover, even though its mid-infrared emission appears to originate from a quite compact region, the corresponding volume and stellar density in the core of \sbs\ are much less extreme than for the super star cluster in NGC~5253.  Finally, the potential for future star formation is substantially more promising for \sbs: the galaxy has a large \HI\ reservoir that spans 66~kpc by 22~kpc (covering both the eastern and western optical components of \sbs), an order of magnitude larger in linear extent than the optical emission. 

%In addition to having an uncommonly large \HI\ envelope, this galaxy has the second lowest metallicity known for an extragalactic system.  Perhaps the material sent into the interstellar medium by stellar outflows and supernova events in the current star formation sites represent the galaxy's first metal enrichment of the interstellar medium.  Moreover, if the bulk of the star formation activity is occurring in at most a handful of super star clusters, then the enrichment should be episodic and stochastic.  The distribution of metals is thus expected to be quite inhomogeneous, with pockets of relatively high metallicity surrounded by a nearly pristine environment.  For example, Thuan, Izotov \& Lipovetsky (1997) have placed an upper limit to the oxygen-to-hydrogen abundance in the neutral gas of \sbs:  $\leq2.7 \times 10^{-7}$, or 3000 times smaller than the oxygen abundance found in Orion and more than three times lower than the oxygen abundance in the \HI\ envelope of I~Zw~18. 

\section{Summary}
The previous work on \sbs\ and our new results suggest that \sbs\ is a galaxy observed early in its formative years, and is likely experiencing its first burst of star formation.  We find that:

\noindent 1. \sbs\ has a very compact distribution of hot dust, which suggests that significant star formation has not occurred for long nor over large spatial scales.  A more extended dust distribution presumably requires more red giant stars processing dust grains for a longer time.

\noindent 2. The rough spectral shape mapped out by the available mid- and far-infrared fluxes suggests that the global distribution of dust is in the form of at least one warm (80~K) component with a mass of $\sim 2400~M_\odot$ and one $\sim 6~M_\odot$ hot (210~K) component; more longer-wavelength data are needed to verify the presence of a substantial cold dust component.

\noindent 3. In spite of the depressed global metallicity, hot dust is one of the energetically dominant features of the galaxy's interstellar medium, resulting in the mid-infrared emission being remarkably comparable to that in the far-infrared.  An important conclusion may be reached here:  young systems that are characterized by a few pockets of active star formation and an overall metal-poor interstellar medium will show elevated mid- to far-infrared ratios.  Perhaps normal characteristics are achieved only after a certain amount of time has passed, after large dust grains have had time to form or build up.  Only then will the more extended, cooler far-infrared dust emission come to dominate the pockets of hot dust emission.
%Here we use the latest photometry from Papaderos et al. (1998) to estimate the blue luminosity assuming $f_\nu (0.44\micron)=4260 \times 10^{-0.4 m_B}$~Jy.
% assuming L_sun_bolometric=3.826e26 W and d=53.3 Mpc:
%L_2.2 = 3e14/2.2mu*6.32e-24*10**(-k/2.5) =5.099e7 L_sun
%L_12.5 = 35e-29*c/12.5mu*4*pi*d^2 where d=53.3e6*3.09e26 m =7.478e8 L_sun
%L_67.3 = 112e-29*c/67.3mu*4*pi*d^2 where d=53.3e6*3.09e26 m =4.44e8 L_sun
%L_0.44 = 4.26e-23*10^(-0.4*16.96)*c/0.44mu*4*pi*d^2 = 4.252e8 L_sun
%but corrected for galactic extinction, L_0.44 = 5.129e8 L_sun
%see /ull/iso_ground/data/computeB... no internal extinction correction to a
% face-on perspective is applied since assumed it's circular.

\noindent 4. The mid-infrared flux coupled with the mid-infrared size and dust temperature indicate that the dust in the star-forming region is optically thin in emission and $A_V<1$~mag.  Since CO molecules likely need to be self-shielding and thus optically thick to be found in appreciable amounts (Meier et al. 2001), it is not surprising that CO emission was not detected for \sbs.

\noindent 5. The galaxy has a bolometric infrared-to-radio ratio that is four times what is expected for normal star-forming galaxies.  This is presumably due to a lack of cosmic rays trapped by interstellar magnetic field lines, as evidenced by the galaxy's negligible nonthermal radio emission.  The dearth of trapped cosmic rays is explained by a combination of two effects: i) the relatively short star-formation history of this galaxy---the starburst is so young that its luminosity is still dominated by extremely massive, luminous stars.  The infrared luminosity output per star is high, but the number of supernova remnants is low, assuming the radio power per supernova remnant is independent of progenitor mass; ii) the compact size of \sbs---dwarf galaxies are known to exhibit low trapping of cosmic rays (Condon, Anderson \& Helou 1991). 

\noindent 6. \sbs\ has an extremely large Br$\gamma$ equivalent width that is indicative of a vigorous star-forming region only 3--6~Myr old.  The distribution of the Br$\gamma$ hydrogen recombination line appears to be more extended than the mid-infrared continuum.  

\noindent 7. The estimated average interstellar heating intensity is $\sim 10,000$ times the intensity found in the solar neighborhood.  Comparable heating intensities are not typically seen globally in galaxies; only \HII\ regions show such elevated heating intensities.

\acknowledgements 
We are happy to acknowledge M. Sauvage for the generous donation of unpublished \ISO\ results.  We are also grateful T. Murphy helped with the LWS observations and the PIFS data reduction, and for the assistance E. Egami provided with the PIFS observing run.  The comments from an anonymous referee helped to improve the text.  The W.M. Keck Observatory is operated as a scientific partnership between the California Institute of Technology, the University of California and the National Aeronautics and Space Administration.  It was made possible by the generous financial support of the W.M. Keck Foundation.  \ISO\ is an ESA project with instruments funded by ESA member states (especially the PI countries: France, Germany, The Netherlands and the United Kingdom) and with the participation of ISAS and NASA.  This research has made use of the NASA/IPAC Extragalactic Database which is operated by JPL/Caltech, under contract with NASA.  B.T.S. is supported by grants from the NSF and NASA.  D.T.F. acknowledges support from NSF grant AST 96-13717 made to the Owens Valley Millimeter Array.

%\newpage

\begin{deluxetable}{lcc}
\small
\def\a{\tablenotemark{a}}
\def\b{\tablenotemark{b}}
\def\c{\tablenotemark{c}}
\def\d{\tablenotemark{d}}
\def\e{\tablenotemark{e}}
\def\f{\tablenotemark{f}}
\def\g{\tablenotemark{g}}
\def\p{$\pm$}
\tablenum{1}
\tablewidth{450pt}
%\tablewidth{0pt}
\tablecaption{Data for \sbs}
\tablehead{
\colhead{Parameter}&\colhead{Value}&\colhead{Source}}
\startdata
optical position (J2000) & $03^{\rm h}37^{\rm m}$44\fs0  $-05\arcdeg02\arcmin38\arcsec$ & NED\nl
%optical position (B1950)& $03^{\rm h}35^{\rm m}$15\fs15 $-05^{\rm d}12^{\rm m}25.9^{\rm s}$ & NED\nl
c$z_\odot$            & 4043\kms                    & NED \nl 
distance\a            & 53.3 Mpc                    & \nl
$m_B$\b               & 16.76 mag                   & Papaderos et al. (1998)\nl
$m_K$\b               & 15.43 mag                   & Vanzi et al. (2000)\nl
$f_\nu(0.44~\micron)$\c& 0.84\p0.02 mJy                   & Papaderos et al. (1998)\nl 
$f_\nu(2.2~\micron)$\c& 0.42\p0.01 mJy                    & Vanzi et al. (2000)\nl 
$f_\nu(12.5~\micron)$ & 35\p6 mJy                   & this work\nl 
$f_\nu(67.3~\micron)$ & 112\p28 mJy                 & this work\nl 
$f$(Br$\gamma)$       & $6\times10^{-18}$~W~m$^{-2}$& this work\nl
$f_\nu(12~\micron)$\d & $<~72$ mJy                  & \IRAS\ \nl
$f_\nu(25~\micron)$\d & $<117$ mJy                  & \IRAS\ \nl
$f_\nu(60~\micron)$\d & $<165$ mJy                  & \IRAS\ \nl
$f_\nu(100~\micron)$\d& $<210$ mJy                  & \IRAS\ \nl
$S_{\rm CO(1-0)}$\d   & $<1.2$ Jy\kms               & this work \nl
$S_{\rm CO(2-1)}$\d   & $<3.0$ Jy\kms               & this work \nl
$f_\nu(1.49$ GHz)     & 0.4\p0.1 mJy                & this work \nl
$L(0.44~\micron)$\e   & $4.3 \times 10^8~L_\odot$   & Papaderos et al. (1998)\nl 
$L(12.5~\micron)$\e   & $7.5 \times 10^8~L_\odot$   & this work\nl 
$L(67.3~\micron)$\e   & $4.4 \times 10^8~L_\odot$   & this work\nl
$M({\rm H}_2)$\f      & $<4.1\times 10^7~M_\odot$   &this work \nl
$M({\rm H}_2)$\g      & $<4.9\times 10^8~M_\odot$   &this work \nl
\enddata
\tablenotetext{a}{Computed in the Local Group reference frame and assuming $H_o=75$~km~s$^{-1}$~Mpc$^{-1}$.}
\tablenotetext{b}{Corrected for Galactic extinction (Schlegel, Finkbeiner \& Davis 1998).}
\tablenotetext{c}{Computed assuming $f_\nu (0.44\micron)=4260 \times 10^{-0.4 m_B}$~Jy and $f_\nu (2.2\micron)=620 \times 10^{-0.4 m_K}$~Jy (Bessel 1979).}
\tablenotetext{d}{$3\sigma$ upper limit.}
\tablenotetext{e}{$L \equiv 4 \pi d^2 \nu f_\nu$ where $\nu f_\nu$ is the power per octave.  The solar luminosity is taken to be $L_\odot=3.826 \times 10^{26}$~W.}
\tablenotetext{f}{Computed assuming a Galactic CO to H$_2$ conversion factor.}
\tablenotetext{g}{Computed assuming a CO to H$_2$ conversion factor equal to twelve times the Galactic value.}
\label{tab:data}
\normalsize
\end{deluxetable}

\begin{deluxetable}{ccc}
\small
\def\a{\tablenotemark{a}}
\def\b{\tablenotemark{b}}
\def\c{\tablenotemark{c}}
\def\d{\tablenotemark{d}}
\def\t{$\times$}
\tablenum{2}
\tablewidth{0pt}
\tablecaption{Sizes of \sbs}
\tablehead{
\colhead{Wavelength}&\colhead{Size}   &\colhead{Size}\\ 
\colhead{}          &\colhead{\arcsec}&\colhead{kpc} }
\startdata
2.166\m\ (Br$\gamma$) & $\sim$1\t1 & 0.26\t0.26 \nl
2.2\m                 & $\sim$2\t1 & 0.52\t0.26 \nl
12.5\m                & 0.31\t0.31 & 0.08\t0.08 \nl
20 cm\a               & $<$6\t6       & 1.55\t1.55 \nl
\enddata
%\tablenotetext{a}{for the region centered on super star clusters \#1 \& \#2 of Thuan et al. (1997)}
\tablenotetext{a}{unresolved}
%\tablenotetext{c}{}
%\tablenotetext{}{1\arcsec\ corresponds to 258 pc}
\label{tab:size}
\normalsize
\end{deluxetable}

%%%%%%%%%%%%%%%%%%%%%%%%%%%%%%%%%%%%%%
%%%%%%%%%%%%%%%%%%%%%%%%%%%%%%%%%%%
%%%%%%%%%%%%%%%%%%%%%%%%%%%%%%%%%%%%%%
%%%%%%%%%%%%%%%%%%%%%%%%%%%%%%%%%%%
%%%%%%%%%%%%%%%%%%%%%%%%%%%%%%%%%%%%%%
%%%%%%%%%%%%%%%%%%%%%%%%%%%%%%%%%%%
%%%%%%%%%%%%%%%%%%%%%%%%%%%%%%%%%%%%%%
%%%%%%%%%%%%%%%%%%%%%%%%%%%%%%%%%%%
%%%%%%%%%%%%%%%%%%%%%%%%%%%%%%%%%%%%%%
%%%%%%%%%%%%%%%%%%%%%%%%%%%%%%%%%%%
%%%%%%%%%%%%%%%%%%%%%%%%%%%%%%%%%%%%%%
%\begin{figure}[!ht]
%\centerline{\psfig{figure=Dale.fig6.ps,width=5.5in,bbllx=185pt,bblly=120pt,bburx=419pt,bbury=668pt,angle=270}}
%\caption[]
%{\ The lefthand panel displays the CO(1-0) map. The rms level is 5.4 mJy beam$^{-1}$, and the contour levels are $1\sigma$ times -4, -3, -2, 2, 3, and 4.  The righthand panel displays the CO(2-1) map. The rms level is 11.8 mJy beam$^{-1}$, and the contour levels are $1\sigma$ times -4, -3, -2, 2, 3, and 4.  Both maps are averaged over 100\kms\ and centered at $03^{\rm h}37^{\rm m}$44\fs1 $-05^{\rm d}02^{\rm m}38^{\rm s}$ (J2000) with north up and east to the left.  The major axis of the beam is oriented at a position angle 170\arcdeg\ east of north and 169\arcdeg\ east of north for the (1-0) and (2-0) maps, respectively.}
%\label{fig:CO_map}
%\end{figure}
%%%%%%%%%%%%%%%%%%%%%%%%%%%%%%%%%%%
%%%%%%%%%%%%%%%%%%%%%%%%%%%%%%%%%%%%%%
%%%%%%%%%%%%%%%%%%%%%%%%%%%%%%%%%%%
\end{document}